\documentclass[conference]{IEEEtran}
\IEEEoverridecommandlockouts
\usepackage{cite}
\usepackage{amsmath,amssymb,amsfonts}
\usepackage{algorithmic}
\usepackage{graphicx}
\usepackage{textcomp}
\usepackage{xcolor}
\usepackage{pgfplots}
\usepackage{cite}
\pgfplotsset{compat=1.5}
\usepackage{pgfplotstable}
\pgfplotsset{grid style={dotted,gray}}
\usepackage{tikz}
\usepackage{xcolor}
\def\BibTeX{{\rm B\kern-.05em{\sc i\kern-.025em b}\kern-.08em
    T\kern-.1667em\lower.7ex\hbox{E}\kern-.125emX}}

\DeclareMathOperator{\pred}{Pred}
\DeclareMathOperator{\layerv}{Layer^v}
\DeclareMathOperator{\layerc}{Layer^c}
\DeclareMathOperator{\messageFN}{Message}
\DeclareMathOperator{\messageC}{Message^c}
\DeclareMathOperator{\messageV}{Message^v}
\DeclareMathOperator{\updateFN}{Update}
\DeclareMathOperator{\aggregateFN}{Aggregate}

\begin{document}

\title{Decoding Quantum LDPC Codes Using\\ Graph Neural Networks
\thanks{Corresponding e-mail: ninkovic@uns.ac.rs}
\thanks{This work has received funding from the European Union's Horizon 2020 research and innovation programme under Grant Agreement number 856967.}
}

\author{\IEEEauthorblockN{Vukan Ninkovic,\IEEEauthorrefmark{1}\IEEEauthorrefmark{2} Ognjen Kundacina, \IEEEauthorrefmark{2}, Dejan Vukobratovic,\IEEEauthorrefmark{1} {Christian H\"{a}ger,\IEEEauthorrefmark{3} and Alexandre Graell i Amat\IEEEauthorrefmark{3}}
\IEEEauthorblockA{\IEEEauthorrefmark{1}Faculty of Technical Sciences, University of Novi Sad, 
Novi Sad, Serbia}
\IEEEauthorblockA{\IEEEauthorrefmark{2}The Institute for Artificial Intelligence Research and Development of Serbia, 
Novi Sad, Serbia}
\IEEEauthorblockA{\IEEEauthorrefmark{3}Department of Electrical Engineering, Chalmers University of Technology, Gothenburg, Sweden
}}}

\maketitle

\begin{abstract}
In this paper, we propose a novel decoding method for Quantum Low-Density Parity-Check (QLDPC) codes based on Graph Neural Networks (GNNs). Similar to the Belief Propagation (BP)-based QLDPC decoders, the proposed GNN-based QLDPC decoder exploits the sparse graph structure of QLDPC codes and can be implemented as a message-passing decoding algorithm. We compare the proposed GNN-based decoding algorithm against selected classes of both conventional and neural–enhanced QLDPC decoding algorithms across several QLDPC code designs. The simulation results demonstrate excellent performance of GNN-based decoders along with their low complexity compared to competing methods.
\end{abstract}


\section{Introduction}

Quantum Error Correction (QEC) is a fundamental technology for future evolution of quantum computing \cite{Devitt}. Mirroring their role in classical error correction, Quantum Low-Density Parity-Check (QLDPC) codes represent one of the most promising QEC classes \cite{MacKay,Postol}. Design and decoding methods for QLDPC codes received increasing attention over the past years \cite{QLDPC,Tillich,Roffe, Camara}. Due to their sparse graph structure, QLDPC codes can be represented using factor graphs and decoded using message-passing Belief Propagation (BP) methods \cite{Tanner,BP}. However, the decoding problem for QLDPC codes is different compared to classical LDPC decoding, since the QLDPC decoder may additionally exploit so-called degeneracy in order to attain better performance. For this reason, a number of recent modifications of the BP decoder for QLDPC codes have been proposed. For example, neural-enhanced BP (NBP) approaches originally proposed for standard LDPC decoding in \cite{NBP} shows excellent performance for decoding QLDPC codes \cite{Kuo,Liu}. In another approach, the BP decoder is enhanced using post-processing based on Ordered Statistics Decoding (OSD), yielding excellent performance albeit under increasing decoding complexity \cite{Panteleev}. A refined BP decoding for quantum codes employing a serial schedule is introduced in \cite{kuo_2020}, where the proposed algorithm outperforms conventional BP while maintaining reduced complexity.

In this paper, we investigate the decoding of QLDPC codes using Graph Neural Networks (GNNs). GNNs bear a number of similarities to BP-based methods \cite{Welling}: they exploit the code's graph structure and can be implemented using node-based calculations and message exchanges across the underlying factor graph of a QLDPC code. Inspired by the similarities of GNNs and BP, in a recent work, we investigated the potential of GNNs in a similar probabilistic inference setup, and demonstrated various performance and implementation benefits of GNNs relative to the BP approach \cite{GNN-SE}. In this paper, we turn our investigation to GNN-based decoding of QLDPC codes. We focus on the design of GNN-based decoders over the QLDPC factor graph and compare the proposed approach against selected classes of both conventional and neural–enhanced algorithms presented in the literature. Although our GNN decoder does not explicitly exploit degeneracy during training, the obtained results show excellent performance compared to state-of-the-art methods. 

GNNs decoders of classical LDPC codes were recently investigated in \cite{GNN-LDPC}. They consider an additive white Gaussian noise (AWGN) channel, where the decoder maps input log-likelihood ratio (LLR) values associated to each codeword bit into an estimated codeword, differently from the considered QLDPC setup where the decoder maps error syndrome values into an estimated error vector. Besides, there are architectural differences in GNNs designed in our work (e.g., we apply attention-based aggregation). Several recent works investigate different approaches to GNN–-based decoding of QLDPC codes \cite{GNN-QLDPC, qldpc_gnn}. The work in \cite{GNN-QLDPC} concentrates on a GNN decoder whose architecture, unlike our GNN architecture, does not build upon the factor graph of the QLDPC code. In [19], the authors introduce a method where GNN layers enhance the classical BP decoding by utilizing knowledge from previous runs. The idea is to address local minima due to trapping sets or short cycles, thus improving the subsequent BP iteration initialization.



\section{QLDPC Codes: Design and Decoding}
\label{Sec_2}

In this section, we review the design and decoding of QLDPC codes. Starting from quantum stabilizer codes, we restrict our attention to QLDPC codes and their decoding using message-passing BP methods.
\vspace{-0.5mm}
\subsection{Quantum Stabilizer Codes}
\label{Stab}
A $[[n,k]]$ quantum stabilizer code, defined by a stabilizer group $\mathcal{S}$, can be understood as a $2^{k}$--dimensional subspace of the $n$--qubit space $\mathbb{C}^{2^{n}}=(\mathbb{C}^2)^{\otimes n}$ \cite{QLDPC}. 
The group $\mathcal{S}$, which has $m=n-k$ independent generators $\langle S_i \rangle_{i=1}^{m}$, $S_i \in \{I, X, Y, Z\}^{n}$ also known as \textit{stabilizer checks}, is a commutative subgroup of the Pauli group $\mathcal{P}_n$ such that $-I^{n}\not\in\mathcal{S}$ \cite{Camara,Panteleev}. As a consequence, the code subspace $\mathcal{C}$ can be defined as the $+1$ eigenspace of $\mathcal{S}$ \cite{Panteleev} according to
\begin{align}
\label{Eq1}
    \mathcal{C}=\{|\psi\rangle \in \mathbb{C}^{2^{n}} \,:\, S_i|\psi\rangle=|\psi\rangle, i=1,\ldots,m\},
\end{align}
where $|\psi\rangle$ represents a quantum state.

Due to a specific quantum phenomenon, known as digitization of an error \cite{Devitt}, quantum errors $\mathbf{e}=\{e_1, \ldots, e_n\}$ can be interpreted as random Pauli operators $P\in\{I, X, Y, Z\}$ which independently act on each qubit (i.e., a Pauli operator can corrupt a quantum state $|\psi\rangle$) \cite{Panteleev,Kuo}. These operators can be represented in a binary form, i.e., a quantum error $\mathbf{e}$ of length $n$ ($n$-qubit error) is defined by a $2n$ bit string $\mathbf{e}_{\textrm{bin}}$ \cite{Panteleev,Liu}
\begin{align}
\label{Eq2}
    \mathbf{e}_{\textrm{bin}}=\prod_{i=1}^{n} X^{x_i}Z^{z_{n+i}} \rightarrow (x_1, \ldots, x_n|z_{n+1}, \ldots, z_{2n}),
\end{align}
where $x_i$ and $z_{n+i}$ are binary exponents of Pauli operators $X$ and $Z$, respectively, that determine the error $e_i$ that affects the $i$-th qubit.

Two Pauli operators, $P$ and $P'$, with binary representations $(\mathbf{x}|\mathbf{z})$ and $(\mathbf{x'}|\mathbf{z'})$, commute if $\langle \mathbf{x},\mathbf{z'}\rangle+\langle \mathbf{z},\mathbf{x'}\rangle =0$ ($\langle \mathbf{a},\mathbf{b}\rangle=\sum_ia_ib_i$) \cite{Panteleev}. The commutation feature is useful for detecting quantum errors. 
By applying Pauli operators on stabilizer generators  $\langle S_i \rangle_{i=1}^{i=m}$, the obtained eigenvalues generate the \textit{error syndrome} $\mathbf{s}$. More precisely, if a Pauli operator commutes/anticommutes with a certain stabilizer, the corresponding eigenvalues will be $+/-1$, respectively. If we map the eigenvalues according to $+1 \rightarrow 0$ and $-1 \rightarrow 1$, $\mathbf{s}=\{s_1, \ldots, s_m\}$ is a binary vector of length $m$, and a nonzero syndrome indicates that an error is detected (as in conventional syndrome decoding) \cite{Kuo}. From \eqref{Eq1}, if an error operator $\mathbf{e}$ is one of the stabilizers, such an error is harmless (if $\mathbf{e}=S_i$, then $\mathbf{e}|\psi\rangle=|\psi\rangle, \forall |\psi\rangle \in \mathcal{C}$) and it is denoted as a \textit{degenerate error} \cite{Roffe}. The main goal of quantum error correction is to infer an error pattern $\hat{\mathbf{e}}$, for which the total error $\mathbf{e}_{\textrm{tot}}=\mathbf{e}+\hat{\mathbf{e}}$ belongs to $\mathcal{S}$ \cite{Liu}. 

A class of stabilizer codes, known as {CSS} quantum codes, is presented by Calderbank, Shor, and Steane \cite{CSS1,CSS2}. This code is constructed by applying a binary mapping to the stabilizer generators $\langle S_i \rangle_{i=1}^{m}$, i.e, its parity-check matrix is constructed as $H=\begin{bmatrix} H_X & \mathbf{0}\\ \mathbf{0} & H_Z \end{bmatrix}$, where each row represents one stabilizer generator \cite{CSS1}. $H_X$ and $H_Z$ represent binary matrices with $n$ columns, and satisfy the orthogonality (or commutativity) condition $H_XH_Z^{T}=\mathbf{0}$ \cite{CSS1}. The error syndrome $\mathbf{s}$ can be obtained as $\mathbf{s}=(s_{\mathbf{x}}, s_{\mathbf{z}})=(H_Z\mathbf{x}, H_X\mathbf{z})$ \cite{Roffe}.  

\vspace{-0.5mm}
\subsection{Quantum LDPC Codes}
\label{QLDPC}

QLDPC codes \cite{MacKay,Postol} are a special case of CSS stabilizer codes whose parity-check matrix $H$ is sparse \cite{Panteleev}, i.e., the number of non-zero elements in its rows and columns is upper bounded by $l$ and $q$, respectively \cite {Roffe}. Similar to their classical counterparts, LDPC codes, a QLDPC stabilizer code $\mathcal{C}$ can be represented using a factor graph $\mathcal{T}$ (also known as a Tanner graph \cite{Tanner}), which is defined by a set of stabilizer generators  $\langle S_i \rangle_{i=1}^{m}$, $\mathcal{T}=\mathcal{T}(\mathcal{S})$ \cite{Panteleev}. Variable nodes in a binary representation $\{v_i\}_{i=1}^{2n}$ correspond to  qubits (columns of $H$), while check nodes $\{c_j\}_{j=1}^{m}$ correspond to stabilizer generators (rows of $H$), with mutual connections if $H_{v_i, c_j}=1$ \cite{Panteleev,MacKay}. The Tanner graph representation opens an opportunity of using different message passing algorithms for QLDPC decoding, as we detail in Sec. \ref{SOTA_alg}.


\vspace{-0.5mm}
\subsubsection{Quantum hypergraph (hgp) product code} 
Firstly introduced by Tillich and Zemor in \cite{Tillich}, this type of code exploits the parity-check matrices of conventional LDPC codes to obtain the matrices $H_X$ and $H_Z$, i.e., a new quantum code can be constructed as a hypergraph (tensor) product of classical codes \cite{QLDPC}. If we suppose that two classical LDPC codes have parity-check matrices $[H_1]_{m_{1}, n_{1}}$ and $[H_2]_{m_{2}, n_{2}}$, then $H_X=[H_1 \otimes I_{m_2, n_2}, I_{m_1, n_1}\otimes H_2^T]$  and $H_Z=[I_{m_1, n_1} \otimes H_2, H_1^T\otimes I_{m_2, n_2}]$ \cite{Liu}. The final QLDPC parity-check matrix $H_{\text{hgp}}$ is constructed as for CSS codes (see Sec.~\ref{Stab}).
\vspace{-0.5mm}
\subsubsection{Quantum bicycle code}
\label{QBC}
In order to fulfill the commutativity condition (see Sec.~\ref{Stab}), a quantum bicycle code \cite{MacKay} is constructed from a self--orthogonal matrix $H_o$ of size $((n-k)/2)\times n$. If we collect all cyclic permutations of a random binary vector $\mathbf{v}$ (with dimension $n/2$) into a matrix $C$, $H_o$ can be obtained as the concatenation of $C$ and its transpose, $H_o=[C, C^T]$ \cite{Liu}. As a last step, $k/2$ random rows from $H_o$ are erased, and by analogy with CSS codes,  $H_{\text{byc}}=\begin{bmatrix} H_o & \mathbf{0}\\ \mathbf{0} & H_o \end{bmatrix}$.

\vspace{-0.5mm}
\subsection{Message-Passing Decoding of QLDPC Codes}
\label{SOTA_alg}

Decoding methods for QLDPC codes are subject of intensive research due to the fact that degenerate errors introduce novel challenges compared to classical LDPC decoders. The Tanner graph representation of QLDPC codes leads to the application of iterative BP message-passing methods \cite{BP}. 

The classical channel model we use in this paper is a pair of binary symmetric channels (BSC) that models $X$ and $Z$ errors as independent and identically distributed events with (physical) error probability $p_f$ \cite{MacKay}. Given $p_f$, the channel affects a QLDPC codeword with a randomly generated error vector $\mathbf{e}_{\textrm{bin}}$, which is in turn transformed into an error syndrome $\mathbf{s}$ that is available to the decoder. Given the syndrome $\mathbf{s}$, the task of the decoder is to produce an estimate of the binary error vector $\hat{\mathbf{e}}$.  

The BP decoder aims to calculate the marginal posterior distribution $p(e_i=1_{i=\{1,\ldots,2n\}}|\mathbf{s})$ for each bit in $\mathbf{e}_{\textrm{bin}}$. This is achieved by passing messages on $\mathcal{T}$ between variable ($v$) and check ($c$) nodes\cite{Liu} according to  
\begin{align}
\label{BP1}
    \mu_{v\rightarrow c}^{\textrm{step: }t+1}=\ell_{v}+\sum_{c'\in\mathcal{N}(v)/c}  \mu_{c'\rightarrow v}^{\textrm{step: }t},
\end{align}
\begin{align}
\label{BP2}
    \mu_{c\rightarrow v}^{\textrm{step: }t+1}=(-1)^{s_c}2\textrm{tanh}^{-1}\prod_{v'\in\mathcal{N}(c)/v} \textrm{tanh}\frac{\mu_{v'\rightarrow c}^{\textrm{step: }t}}{2},
\end{align}
where $\ell_{v}$ represents the (prior) log-likelihood ratio (LLR) values associated with $e_{v}$, $s_{c}$ is a syndrome value $\mathbf{s}$ on the check node $c$ (position $c$), and $\mathcal{N}(a)/b$ refers to all neighbours of $a$ except $b$. After a given maximum number of iterations (steps), the (posterior) LLR value of each $\{e_i\}_{i=1}^{i=2n}$ is calculated as
\begin{align}
\label{BP3}
    \mu_{v}=\ell_{v}+\sum_{c\in\mathcal{N}(v)} \mu_{c\rightarrow v}^{\textrm{final step}}. 
\end{align}
For decoding QLDPC codes, the BP decoder is affected not only by short cycles in the Tanner graph, but also the fact that it does not exploit degeneracy \cite{Roffe}, \cite{Panteleev}. In order to solve this problem, OSD \cite{OSD} is proposed in a quantum setting as a BP post-processing step \cite{Panteleev, Roffe}, which is used if BP algorithm fails to converge. The main idea of BP-OSD decoding is that soft decisions of BP decoder can be used as an input to OSD, i.e., they define a basis set with the most probable flipped bits (for more details, see \cite{Roffe} and \cite{Panteleev}). 

Since BP-OSD has high complexity \cite{OSD}, the authors in \cite{Liu} proposed, by analogy with classical codes \cite{NBP}, neural-enhanced decoder for quantum codes, where learnable parameters are introduced inside the conventional BP equations (Eqs. \eqref{BP1}-\eqref{BP3}). This approach is known as NBP and can be described using the the modified BP equations \cite{Liu} 
\begin{align}
    \mu_{v\rightarrow c}^{\textrm{step: }t+1}=\ell_{v}b_{v}^{\textrm{step: }t}+\sum_{c'\in\mathcal{N}(v)/c}  \mu_{c'\rightarrow v}^{\textrm{step: }t}w_{c'v}^{\textrm{step: }t}, 
\end{align}
\begin{align}
    \textrm{af}(\mu_{c\rightarrow v}^{\textrm{step: }t+1})=
    i\pi s_{c} + \sum_{v'\in\mathcal{N}(c)/v} \gamma(\mu_{v'\rightarrow c}^{\textrm{step: }t}),
\end{align}
\begin{align}
    \mu_{v}=\ell_{v}b_{v}^{\textrm{final step}}+\sum_{c\in\mathcal{N}(v)} \mu_{c\rightarrow v}^{\textrm{final step}}w_{cv}^{\textrm{final step}},
\end{align}
where $w_{c'v}$, $w_{cv}$ and $b_v$ represent learnable weights and biases, respectively, while $\gamma(x)=\log[\tanh(\frac{x}{2})]$ denotes a nonlinear activation function \cite{Liu}. The NBP decoder is trained by minimizing an appropriate loss function and applying stochastic gradient descent to adjust all trainable parameters. 

\section{GNN-Based Decoding of QLDPC Codes } \label{gnn_qldpc}

In this work, we apply GNNs as a decoder for QLDPC codes. 
We consider a class of message-passing GNNs that exchange messages among neighboring nodes of the underlying graph resembling the BP message passing process \cite{Welling}. We present a short review of GNNs, and proceed with describing the GNN-based QLDPC decoders proposed in this paper.

\begin{figure*}[t]
	\centering
\includegraphics[width=0.75\linewidth]{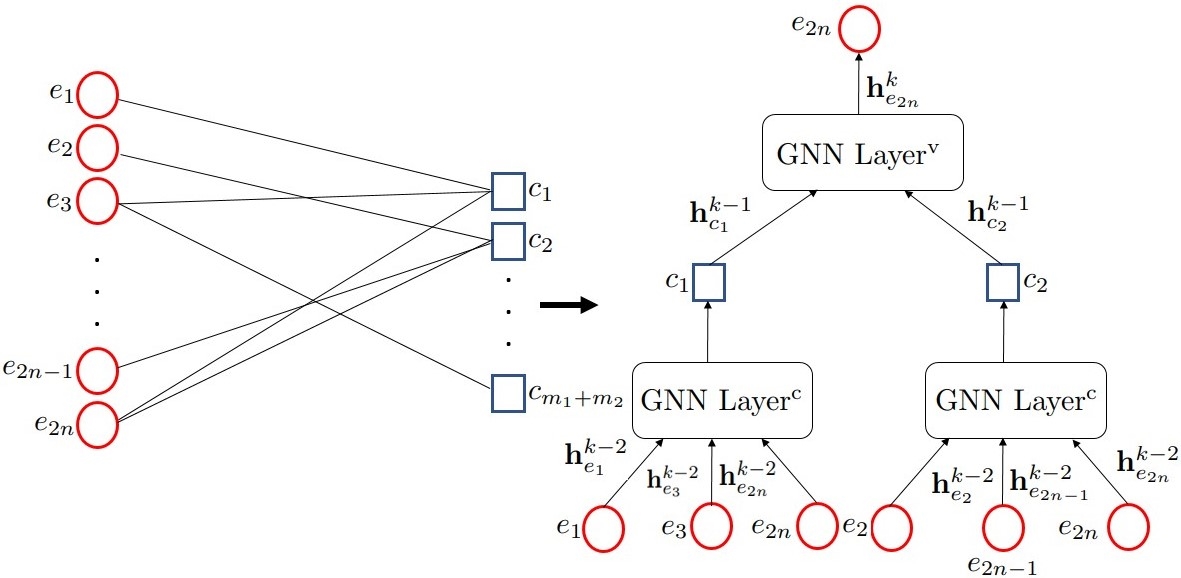}
	\vspace*{-3mm}
	\caption{QLDPC code parity-check matrix $H$ represented as a Tanner graph and its respective underlying GNN structure (with message flow).}
	\label{Fig_GNN}
\end{figure*}

\vspace{-0.5mm}
\subsection{Graph Neural Networks}
\label{GNN_brief}

GNNs are a powerful class of deep learning methods uniquely suited to capture the complex relationships and patterns that exist within graphs. Two main types of GNNs are spectral and spatial (that are also known as message-passing GNNs). Spectral GNNs are based on trainable graph convolutions in the Fourier domain, and therefore cannot generalize to new graphs, are limited to undirected graphs, and can be computationally expensive \cite{spectralGNN}. Message-passing GNNs do not suffer from these issues since they do not rely on matrix-based Fourier transforms. They perform recursive neighbourhood aggregation, also known as message passing, over the local subsets of graph-structured inputs to create a meaningful representation of the connected pieces of data \cite{pmlr-v70-gilmer17a}. In other words, they act as trainable local graph filters which have a goal of transforming the inputs from each node and its connections to a higher dimensional space, resulting in an $s$-dimensional vector embedding $\mathbf h \in \mathbb {R}^{s}$ per node. The message-passing GNN layer, which represents one message passing iteration, combines multiple trainable functions commonly implemented as neural networks.

To illustrate how message-passing GNNs operate, we consider a binary classification problem (note that the QLDPC decoding problem, i.e., the recovery of binary error vector, falls into this category). The task is to categorize each node in the graph into one of two classes, based on input data for each node, and connectivity between the nodes. The prediction process consists of several iterations of message calculations between pairs of connected nodes, which are then combined to produce useful vector representations of nodes. These representations can be used to perform the classification or any other prediction task. The message function $\messageFN(\cdot | \theta^{\messageFN}): \mathbb {R}^{2s} \mapsto \mathbb {R}^{u}$ calculates the message $\mathbf m_{i\rightarrow j} \in \mathbb {R}^{u}$ between two node embeddings, $\mathbf h_i$ and $\mathbf h_j$. The aggregation function $\aggregateFN(\cdot | \theta^{\aggregateFN}): \mathbb {R}^{\textrm{deg}(j) \cdot u} \mapsto \mathbb {R}^{u}$ combines the arriving messages, and outputs the aggregated messages $\mathbf {m}_j \in \mathbb {R}^{u}$ for each node $j$. At the end of one message-passing iteration, the aggregated messages are fed into the update function $\updateFN(\cdot | \theta^{\updateFN}): \mathbb {R}^{u} \mapsto \mathbb {R}^{s}$, to calculate the update of each node's embedding.

The node embedding values are initialized with the node's input feature vector transformed to the $s$-dimensional space, after which the message passing repeats $K$ times, with $K$ being a hyperparameter known as the number of GNN layers. The overall message passing process is described as
\begin{equation}
    \begin{gathered}
        \mathbf {m}_{i\rightarrow j}^{k-1} = \messageFN( \mathbf {h}_i^{k-1}, \mathbf {h}_j^{k-1} | \theta^{\messageFN}),\\
        \mathbf {m}_j^{k-1} = \aggregateFN(\{\mathbf {m}_{i\rightarrow j}^{k-1} | i \in \mathcal{N}_j\} | \theta^{\aggregateFN}),\\
        \mathbf {h}_j^k = \updateFN(\mathbf {m}_j^{k-1}, \mathbf {h}_j^{k-1}) | \theta^{\updateFN}),\\
        k \in \{1,\dots,K\},
    \end{gathered}
    \label{gnn_equations}
\end{equation}
where $\mathcal{N}_j$ denotes the $1$-hop neighbourhood of the node $j$. At the end of the message passing process, the final node embeddings $\mathbf {h}_j^{K}$ are fed through additional neural network layers to produce the final predictions. For binary classification tasks, this typically involves passing the embeddings through additional feed-forward neural network layers, which produces a set of values called logits, i.e., the strength of evidence for each class. The logits are finally passed through a sigmoid function to obtain probability estimates of each node belonging to each class. The prediction process can be described as
\begin{equation}
\begin{gathered}
\mathbf{z}_j = \text{FeedForward}(\mathbf{h}_j^{K} | \theta^{\text{FeedForward}}),\\
\mathbf{\hat{y}}_j = \text{Sigmoid}(\mathbf{z}_j),
\end{gathered}
\label{classification_prediction}
\end{equation}
where $\mathbf{z}_j$ are the logits for the node $j$, obtained by passing its final embedding $\mathbf{h}_j^{K}$ through an additional linear layer with parameters $\theta^{\text{Linear}}$, and $\mathbf{\hat{y}}_j$ are the estimated binary class probability. In Sec.~\ref{gnn_qldpc_subsec}, we will refer to the composition of these two functions given in \eqref{classification_prediction} as $\pred(\cdot|\theta^{\pred})$.

During training, the model makes predictions for a subset of nodes called a mini-batch, and an individual loss for each node is calculated based on these predictions using binary cross entropy (BCE). The total loss $L(\theta)$ is obtained by averaging the individual losses for the nodes in the mini-batch as
\begin{equation}
\label{BCE}
L(\theta) = -\frac{1}{B} \sum_{i=1}^{B} [y_i^{\text{label}} \log(\hat{y}_i) + (1 - y_i^{\text{label}}) \log(1 - \hat{y}_i)],
\end{equation}
where $B$ is the number of nodes in the mini-batch, $\theta$ denotes all the trainable parameters of the GNN model, $\hat{y}_i$ is the predicted probability for a node $i$, and $y_i^{\text{label}}$ is the corresponding ground-truth label (0 or 1). The training objective is to minimize this loss by adjusting the trainable parameters in the entire GNN computational graph, typically using some gradient descent-based algorithm.

\vspace{-0.5mm}
\subsection{GNN-based Decoding of QLDPC Codes } \label{gnn_qldpc_subsec}

The underlying graph structure of QLDPC code (based on a parity-check matrix $H$ and represented by the Tanner graph $\mathcal{T}$ in Sec.~\ref{QLDPC}) is suitable for application of message-passing GNNs. A bipartite graph $\mathcal{T}$ is mapped onto a GNN architecture, where each variable node $\{v_i\}_{i=1}^{i=2n}$ corresponds to an element of the error operator $\mathbf{e}_{\textrm{bin}}$, elements of the syndrome vector $\mathbf{s}$ correspond to $m=m_1+m_2$ check (factor) nodes  $\{c_j\}_{j=1}^{j=m}$, and connections between variable and check nodes are preserved from $\mathcal{T}$, as illustrated on the left-hand side of  Fig.~\ref{Fig_GNN}. 

As described in Sec. \ref{SOTA_alg}, we model quantum errors as $2n$-bit strings $\mathbf{e}_{\textrm{bin}}$ affected by bit flips modelled as a pair of BSCs of physical error rate $p_{f}$ \cite{MacKay}. We use a binary index encoding for the variable node input features, which, compared to one-hot encoding, offers significant advantages, including a reduced number of input neurons, trainable parameters, and improved training/inference time. The GNN check node exploits the input error syndrome $\mathbf{s}$ (Sec.~\ref{Stab}), i.e., the input to each $\{c_j\}_{j=1}^{j=m_1+m_2}$ is the corresponding binary value of the vector $\mathbf{s}$. 

Given that the graph for the proposed GNN consists of two distinct node types, we utilize separate GNN layers when aggregating messages into each of these node types. These layers, labeled as $\layerc(\cdot|\theta^{\layerc}): \mathbb {R}^{\textrm{deg}(f) \cdot s} \mapsto \mathbb {R}^{s}$ and $\layerv(\cdot|\theta^{\layerv}): \mathbb {R}^{\textrm{deg}(v) \cdot s} \mapsto \mathbb {R}^{s}$, possess their own trainable parameters $\theta^{\layerc}$ and $\theta^{\layerv}$. This enables independent learning of their message, aggregation, and update functions. Variable-to-check and check-to-variable node messages are calculated using the corresponding message functions in $\layerc(\cdot|\theta^{\layerc})$ and $\layerv(\cdot|\theta^{\layerv})$ as:
\begin{equation}
    \begin{gathered}
        \mathbf {m}_{v\rightarrow c}^{k-1} = \messageC( \mathbf {h}_v^{k-1}, \mathbf {h}_c^{k-1} | \theta^{\messageC})\\
        \mathbf {m}_{c\rightarrow v}^{k-1} = \messageV( \mathbf {h}_c^{k-1}, \mathbf {h}_v^{k-1} | \theta^{\messageV})\\
        k \in \{1,\dots,K\}.
    \end{gathered}
    \label{gnn_messages}
\end{equation}

Additionally, we apply a neural network $\pred(\cdot|\theta^{\pred}): \mathbb {R}^{s} \mapsto \mathbb {R}$ to the final variable node embeddings $\mathbf h^K$, generating predictions of true error vector $\mathbf{e_{\textrm{bin}}}$. The whole process, illustrated as a computational graph for a single iteration (i.e., GNN layer), is illustrated in the right-hand side of Fig. \ref{Fig_GNN}.

The main goal of conventional QEC decoding is to find an appropriate error pattern (in binary representation), which belongs to a normalizer group spanned by  $H^{\perp}$ (Sec. \ref{Stab}), i.e., the following condition must be satisfied \cite{Liu}: 
\begin{align}\label{Loss}
    H^{\perp}M\mathbf{e}_{\textrm{tot}_{\textrm{bin}}}=\boldsymbol{0},
\end{align}
where $M=\left[ {\begin{array}{cc}
    0 & I_n\\
    I_n & 0\\
  \end{array} } \right]$ and $\mathbf{e}_{\textrm{tot}_{\textrm{bin}}}=\mathbf{e_{\textrm{bin}}}+\hat{\mathbf{e}}_{\textrm{bin}}$.

In order to obtain the estimated error operator $\hat{\mathbf{e}}_{\textrm{bin}}$, the GNN output produces estimate of the probability $p(\hat{e}_i=1|\boldsymbol{s}), i \in \{1,2,...,2n\}$, which is exploited for each of $2n$ values in $\hat{\mathbf{e}}_{\textrm{bin}}$. In the proposed approach, we train the GNN based on the BCE loss function in Eq.~\eqref{BCE} between $\mathbf{e}_{\textrm{bin}}$ (which is used as a label in the training process) and its estimation $\hat{\mathbf{e}}_{\textrm{bin}}$, i.e., the main goal of the GNN-based decoding algorithm is to estimate the true error pattern $\mathbf{e}_{\textrm{bin}}$. 
\vspace{-3mm}
\begin{table}[h]
\caption{List of GNN hyperparameters.}
\vspace{-3mm}
\label{tbl_hyperparameters}
\begin{center}
    \begin{tabular}{ | l | c | }
        \hline
        \textbf{Hyperparameters} & \textbf{Values} \\ 
        \hline
        Number of GNN layers $K$ & $6$ \\
        \hline
        Node embedding size $s$ & $128$ \\
        \hline
        Learning rate & $4\times 10^{-4}$ \\
        \hline
        Minibatch size & $32$ \\
        \hline
        Gradient clipping value & $0.5$ \\
        \hline
        Optimizer & Adam \\
        \hline
    \end{tabular}
\end{center}
\vspace{-6mm}
\end{table}

\section{Performance Evaluation}
\label{Sec_4}

\begin{figure}[t]
	\begin{tikzpicture}
  	\begin{loglogaxis}[width=1\columnwidth, height=6cm, 
	legend style={at={(0.82,0.39)}, anchor= north,font=\scriptsize, legend style={nodes={scale=0.99, transform shape}}},
   	legend cell align={left},
	legend columns=1,   	 
   	x tick label style={/pgf/number format/.cd,fixed,
   	 precision=1, /tikz/.cd},
   	y tick label style={/pgf/number format/.cd,fixed, precision=1, /tikz/.cd},
   	xlabel={Physical error rate ($p_f$)},
   	ylabel={Logical error rate},
   	label style={font=\footnotesize},
   	grid=major,   	
   	xmin =0.0004, xmax = 0.01,
   	ymin=0.00007, ymax=0.1,
   	line width=0.85pt,
   	tick label style={font=\footnotesize},]
    \addplot[blue, mark=square*] 
   	table [x={x}, y={y}] {./Figs/HGP/bp.txt}; 
   	\addlegendentry{BP}
    \addplot[cyan, mark=diamond] 
   	table [x={x}, y={y}] {./Figs/HGP/bp_osd.txt}; 
   	\addlegendentry{BP+OSD}
    \addplot[green, mark=triangle] 
   	table [x={x}, y={y}] {./Figs/HGP/nbp.txt}; 
   	\addlegendentry{NBP}
    \addplot[red, mark=x] 
   	table [x={x}, y={y}] {./Figs/HGP/gnn.txt}; 
   	\addlegendentry{GNN}
    \end{loglogaxis}
	\end{tikzpicture}
	\vspace*{-8.5mm}
	\caption{Performance comparison of the proposed GNN-based QLDPC decoder versus other decoders presented in the literature for different physical error rates $p_f$ (HPG code, $[[n, k]]=[[129, 28]]$). (Note: NBP results are, under identical setting, adopted from \cite{Liu}.)}
	\label{Fig_hgp}
\end{figure}
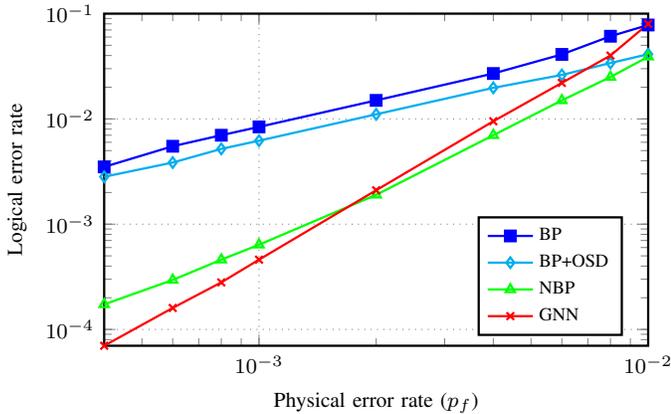

In this section, we compare the proposed GNN-based QEC decoding algorithm against selected classes of both conventional and neural--enhanced algorithms presented in the literature. In all conducted experiments, parameters for benchmark algorithms are taken from \cite{Liu} and \cite{Roffe} i.e., the maximum number of iterations for BP is set to 12, while the OSD order $o$ is set to $o=0$ (quantum bicycle codes) and $o=4$ (quantum hgp and bicycle codes), respectively. Additionally, in all the experiments, we used the same GNN architecture. We employ two-layer feed-forward neural networks for message functions and for the $\pred(\cdot|\theta^{\pred})$ function, a single-layer gated recurrent unit (GRU) neural network for update functions, and the attention mechanism in the aggregation function. For all feed-forward neural network layers, we utilize ReLU activation functions, except for the final layer of $\pred(\cdot|\theta^{\pred})$, where we used sigmoid activation function. The other GNN model and training hyperparameters presented in Table~\ref{tbl_hyperparameters}.

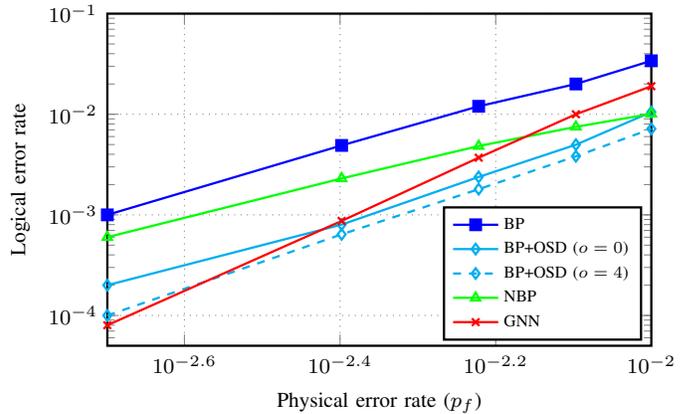
\begin{figure}[t]
	\begin{tikzpicture}
  	\begin{loglogaxis}[width=1\columnwidth, height=6cm, 
	legend style={at={(0.8,0.42)}, anchor= north,font=\scriptsize, legend style={nodes={scale=0.9, transform shape}}},
   	legend cell align={left},
	legend columns=1,   	 
   	x tick label style={/pgf/number format/.cd,fixed,
   	 precision=1, /tikz/.cd},
   	y tick label style={/pgf/number format/.cd,fixed, precision=1, /tikz/.cd},
   	xlabel={Physical error rate ($p_f$)},
   	ylabel={Logical error rate},
   	label style={font=\footnotesize},
   	grid=major,   	
   	xmin =0.002, xmax = 0.01,
   	ymin=0.00005, ymax=0.1,
   	line width=0.85pt,
   	tick label style={font=\footnotesize},]
    \addplot[blue, mark=square*] 
   	table [x={x}, y={y}] {./Figs/Byc/bp.txt}; 
   	\addlegendentry{BP}
    \addplot[cyan, mark=diamond] 
   	table [x={x}, y={y}] {./Figs/Byc/osd_0.txt};
    \addlegendentry{BP+OSD ($o=0$)}
    \addplot[dashed, cyan, mark=diamond, mark options={solid}] 
   	table [x={x}, y={y}] {./Figs/Byc/osd.txt};
   	\addlegendentry{BP+OSD ($o=4$)}
    \addplot[green, mark=triangle] 
   	table [x={x}, y={y}] {./Figs/Byc/nbp.txt}; 
   	\addlegendentry{NBP}
    \addplot[red, mark=x] 
   	table [x={x}, y={y}] {./Figs/Byc/gnn.txt}; 
   	\addlegendentry{GNN}
    \end{loglogaxis}
	\end{tikzpicture}
	\vspace*{-8.5mm}
	\caption{Performance comparison of the proposed GNN-based QLDPC decoder versus other decoders presented in the literature for different physical error rates $p_f$ (Bicycle code, $[[n, k]]=[[256, 32]]$). (Note: NBP results are, under identical setting, adopted from \cite{Liu}.)}
	\label{Fig_bic}
\end{figure}

\vspace{-0.5mm}
\subsection{Data Set Generation}
The data set consists of $5\times10^3$ (for training) and $10^6$ (for testing) ($\mathbf{s}$, $\mathbf{e}_{\textrm{bin}}$) pairs, where the obtained syndrome ($H\mathbf{e}_{\textrm{bin}}=\mathbf{s}$) represents the input to the GNN check nodes, while the true binary error vector defines a label (Sec.~\ref{gnn_qldpc_subsec}). In all experiments, the training procedure is conducted with a physical error rate $p_f=0.01$, while testing is done across a range of different $p_f$ values. 

As a consequence of small physical error rate in the training phase, randomly created data set would be unbalanced containing large fraction of all-zero $\mathbf{e}_{\textrm{bin}}$ vectors. In order to overcome this problem, we design a training data set in semi-controlled manner, where the first $2n+1$ $\mathbf{e}_{\textrm{bin}}$ vectors: the all-zero vector and the set of $2n$ vectors that define standard basis of the field $F^{2n}$ are automatically included in the data set. The remaining $\mathbf{e}_{\textrm{bin}}$ examples in the training set are obtained as follows: for each $\mathbf{e}_{\textrm{bin}}$, we first choose the number of ones (2 or more, up to $2n$) from an appropriate re-normalized distribution, and distribute the selected number of ones randomly across $\mathbf{e}_{\textrm{bin}}$. For the test data set, this procedure is not used.  

\vspace{-0.5mm}
\subsection{Quantum Hypergraph--Product (hgp) Code} We test our approach on the hgp code with parameters $[[n,k]]=[[129, 28]]$ presented in \cite{Liu}, where conventional $[7, 4, 3]$ and $[15, 7, 5]$ BCH codes are used as $H_1$ and $H_2$ in the QLDPC construction of $H_{\text{hgp}}$ (see Sec.~\ref{QLDPC}). From Fig.~\ref{Fig_hgp}, we can observe that, although the presented GNN-based approach does not explicitly exploit degeneracy during training, it significantly outperforms conventional algorithms, such as BP and BP+OSD, simultaneously showing comparable or better performance than NBP. As expected, the conventional BP decoder fails to converge due to the short cycles present in the graph. On the other hand, in \cite{Roffe}, it is shown that for random QLDPC codes with shorter blocklengths, OSD post-processing does not lead to significant gains, which is the case in our example. Moreover, the complexity of the OSD scales less favourably than of the GNN-based approach (see Sec.~\ref{Complexity}).    

\vspace{-0.5mm}
\subsection{Quantum Bicycle Code}

In Fig. \ref{Fig_bic}, we compare the performance of the presented GNN-based decoding algorithm with other decoding algorithms from the literature (BP, BP+OSD, NBP) for a quantum bicycle code. The code parameters are taken from \cite{Liu} ($[[n,k]]=[[256,32]]$), i.e., $H_{\text{byc}}$ is constructed from a sparse matrix $H_o$, where the weight (number of ones) of the vector $\mathbf{v}$ is 8 (Sec.~\ref{QBC}). From Fig. \ref{Fig_bic}, we note that the GNN approach significantly outperforms BP and NBP. 
We note a slight performance degradation compared to BP+OSD for larger physical error rates, although the presented GNN-based approach has a steeper slope. This is expected due to the above-mentioned behavior of the BP+OSD decoder with the increase of the code blocklength. In order to reduce the complexity of the BP+OSD decoder, we first implement the \textit{OSD-0 algorithm} from \cite{Roffe}, where the OSD order is $o=0$, and, as we can see from Fig.~\ref{Fig_bic} there is a notable performance degradation compared to the  higher-order OSD ($o=4$) which, on the other hand, imposes significantly higher complexity (see Sec.~\ref{Complexity}).

\vspace{-0.5mm}
\subsection{Comment on Decoder Complexity Scaling}
\label{Complexity}

We present a brief discussion on the complexity of the presented algorithms, both conventional and neural enhanced. According to \cite{OSD}, complexity of the OSD algorithm is $\mathcal{O}(k^{o})$, i.e., it increases with the order $o$ in the exponent (we plot results for BP+OSD with OSD order $o=0$ in Fig. \ref{Fig_bic}). On the other hand, as mentioned in Sec. \ref{QLDPC}, the check node degree in the QLDPC parity-check matrix ($H_{\text{hgp}}$/$H_{\text{byc}}$) is upper bounded by a constant value $l$. As a consequence, the overall complexity of BP and NBP algorithms is linear in a code length n ($\mathcal{O}(n)$) \cite{Smith}, where NBP complexity can be further tuned by applying a pruning approach \cite{Buchberger}. For the same reason (bounded check node degree), GNN-based decoding during inference phase will also have linear complexity \cite{GNN-LDPC}. Note that GNNs typically require a smaller number of iterations (layers) to achieve similar performance as BP \cite{GNN-SE}, however, the layer complexity depends on its architecture. For example, for bicycle codes (Fig.~\ref{Fig_bic}), the GNN-based decoder achieves comparable results with BP+OSD decoder ($o=4$) under linear $O(n)$ vs $O(k^4)$ complexity scaling. A detailed analysis, which includes implementation challenges and number of FLOPs calculation, is out of the scope of this work, and will be addressed in future work. 

\section{Conclusion}
\label{Sec_5}

In this work, we proposed, designed and evaluated a GNN-based decoder for QLDPC codes. The GNN decoder exploits the underlying QLDPC code factor graph, operates by passing messages over the factor graph and shares the same linear complexity scaling as the BP decoder. Simulation results demonstrate excellent performance of the GNN-based decoder tested across several QLDCP code designs and against relevant state-of-the-art QLDPC decoders from the recent literature.

\vspace{12pt}

\end{document}